# Group characterizable entropy functions

Terence H. Chan
Institute for Telecommunications Research
University of South Australia, Australia
terence.chan@unisa.edu.au

*Abstract*— This paper studies properties of entropy functions that are induced by groups and subgroups. We showed that many information theoretic properties of those group induced entropy functions also have corresponding group theoretic interpretations. Then we propose an extension method to find outer bound for these group induced entropy functions.

## I. INTRODUCTION

The determination of capacity regions in communication systems (such as a channel or a network) is a classical information theory problem. Due to lack of tools, capacity regions are usually hard to determine except in some simple cases. Some examples of the available tools include Shannon and non–Shannon Information inequalities and LP bound.

Take network multicast problem as an example. Consider data generated at multiple sources to be multicast to various receivers in the network. Each source is associated with a random variable denoting the random message generated at that source. Similarly, each edge/channel is associated with a random variable denoting the random message transmitted along the edge. Clearly, how these random variables relate to each other is not arbitrary but must depend on the transmission scheme and the underlying network topology. Specifically, they must meet the following criteria: (1) random variables associated with sources are independent to each other, (2) random variable with an outgoing edge of a node $v$ must be a deterministic function of the set of random variables associated with sources can channels directing towards $v$, and (3) if the data generated at a source is destined to a node, then the random variable with that source must be a function of the random variables associated with incoming messages and sources available to that node.

The capacity region for the multicast problem is essentially equivalent to determining the set of all entropy functions for sets of random variables meeting the above criteria. By using various information inequalities and LP bound, we obtain an outer bound for the capacity region by finding outer bounds for the set of entropy functions.

If there is a restriction on the class of network codes (say, linear network codes) that can be used, the set of entropy functions induced by the variables may meet additional requirements. For example, in [1], it was shown that if an abelian network codes (including linear network codes and R–module codes as special cases) are used, the set of random variables can then be "characterized" by abelian groups and consequently, its entropy function satisfies additional requirement such as the Ingleton's inequality. Moreover, the capacity region for linear network codes can be determined similarly by identifying the set of entropy functions induced by the abelian groups.

Unfortunately, the above characterization for the capacity region usually cannot be implemented in practice because the difficulty of determining the set of entropy functions meeting the required criteria. LP bound gives outer bounds for capacity region by replacing the set of entropy functions by the set of polymatroidal functions. Clearly, if one can tighten the bound for the set of entropy functions (or those induced by abelian groups), then one can get a tighter capacity region bound.

The organization of the paper is as follows. In Section 2, we describe how groups and subgroups can induce random variables and demonstrate that many information theoretic properties of these group induced random variables have nice group theoretic interpretations. Section 3 focuses on the properties of those random variables induced by abelian groups including vector spaces. These abelian group induced random variables is of special interests because of its relation to linear network code capacity region. And finally in Section 4, we propose a method to outer bound the set of entropy functions by using a method called "extension".

## II. GROUPS, RANDOM VARIABLES AND ENTROPIES

Random variables (and their corresponding entropy functions) induced/characterized by groups and subgroups have many nice properties. For example, [2] proved that an unconstrained information inequality is valid if and only if it holds for all random variables induced by groups and subgroups. Through this relation, tools in group theory can be used to prove information inequalities, or vice versa.

While [2] proved the equivalence of information theoretic and group theoretic inequalities, this paper focuses on the properties of entropy functions, especially those generated from abelian groups and subgroups.

Let $\mathsf{N} = \{1, \cdots, n\}$ and $X_1, X_2, \cdots, X_n$ be $n$ jointly distributed discrete random variables. For any nonempty subset $\alpha$ of $\mathsf{N}$, $X_\alpha$ denotes the joint random variable $(X_i : i \in \alpha)$ whose entropy is denoted by $H(X_\alpha)$. Let $\mathcal{F}_n$ be the set of all real functions defined on the collection of nonempty subsets of $\mathsf{N}$. Hence, it is a $(2^n - 1)$-dimensional real Euclidean space [7].

**Definition 1.** *Let* $\mathbf{g} \in \mathcal{F}_n$. *Then* $\mathbf{g}$ *is an entropy function (or is called entropic) if there exists a set of random variables* $X_1, X_2, \cdots, X_n$ *such that* $\mathbf{g}(\alpha) = H(X_\alpha)$ *for any nonempty subset* $\alpha$ *of* $\mathsf{N}$.

Let $\overline{\Gamma}_n^*$ be the set of all entropy functions and hence is a subset of $\mathcal{F}_n$. It plays an important role in information theory; however, it also has a very complex structure. In [7], it was proved that $\overline{\Gamma}_n^*$, the closure of $\Gamma_n^*$, is a closed and convex cone, and thus is much more manageable than $\Gamma_n^*$. In fact, for many applications, it is sufficient to consider $\overline{\Gamma}_n^*$ instead of $\Gamma_n^*$. However, except for $n \leq 3$, a complete characterization for the set $\overline{\Gamma}_n^*$ is still absent.

The main result on the relations between groups, random variables and entropies is the following theorem proved in [2].

**Theorem 1.** *Let $G$ be a finite group and $G_1, G_2, \cdots, G_n$ be its subgroups. Then there exists a set of random variables $X_1, X_2, \cdots, X_n$ (said to be induced/characterized by $(G, G_1, G_2, \cdots, G_n)$) such that $H(X_\alpha) = \log |G|/|\bigcap_{i \in \alpha} G_i|$ for any nonempty subset $\alpha$ of $\mathsf{N}$.*

**Proof:** Suppose $U$ is a random variable having a uniform probability distribution over $G$. For each $i \in \mathsf{N}$, the subgroup $G_i$ partitions the group $G$ into $\log |G|/|G_i|$'s left cosets of $G_i$ in $G$. Then we define the random variable $X_i$ as a function of $U$ such that $X_i$ is the index for the left coset of $G_i$ in $G$ that contains $U$. It was then shown in [2] that $H(X_\alpha) = \log |G|/|\bigcap_{i \in \alpha} G_i|$ for any nonempty subset $\alpha$ of $\mathsf{N}$. ∎

Many nice information theoretic properties of random variables induced by groups have group theoretic interpretations. Some examples are given as follows.

**Proposition 1** (Closed in addition). *Let $\mathbf{g}_1$ and $\mathbf{g}_2$ be group characterizable (i.e., $\mathbf{g}_1$ and $\mathbf{g}_2$ are entropy functions for random variables induced by groups). Then $\mathbf{g}_1 + \mathbf{g}_2$ is also group characterizable.*

**Proposition 2** (Conditioning). *Let $\mathbf{g} \in \mathcal{F}_n$ be group characterizable. For any non-empty subset $\beta$ of $\mathsf{N}$, define a function $\hat{\mathbf{g}}_\beta \in \mathcal{F}_n$ such that $\hat{\mathbf{g}}_\beta(\alpha) = \mathbf{g}(\alpha \cup \beta) - \mathbf{g}(\beta)$ for all subset $\alpha$ of $\mathsf{N}$. Then $\hat{\mathbf{g}}_\beta$ is group characterizable.*

Let $(X_1, X_2, \cdots X_n)$ be a set of $n$ random variables and $\mathbf{g}$ be its entropy function. The proposition said that if $(X_1, X_2, \cdots X_n)$ is group characterizable, then the function $\hat{\mathbf{g}}_\beta(\alpha) \triangleq H(X_\alpha | X_\beta)$ for any non-empty subset $\alpha$ of $\mathsf{N}$ is also group characterizable.

**Proposition 3** (Functional dependency). *Let $(X_1, X_2, \cdots X_n)$ be a set of random variables induced by a group characterization $(G, G_1, G_2, \cdots, G_n)$. For any element $i$ and subset $\alpha$ of $\mathsf{N}$, we have $H(X_i | X_j : j \in \alpha) = 0$ if and only if $\bigcap_{j \in \alpha} G_j$ is a subgroup of $G_i$.*

**Remark:** Proposition 3 is very useful in determining capacity region for network codes. In [1], linear (and abelian) network codes are interpreted as group characterizations for the set of random message variables transmitted along the network links. By the proposition, the functional dependency between outgoing message and the incoming messages of a node can be rephrased as that the intersection of subgroups associated with incoming edges to the node is contained in the subgroup associated with the outgoing edge.

**Proposition 4** (Independence). *Let $X_1, \cdots, X_n$ be induced by the group characterization $(G, G_1, \cdots, G_n)$. Then $X_1, \cdots, X_n$ are independent (i.e., $H(X_1, \cdots, X_n) = \sum_{i=1}^n H(X_i)$ if and only if $\prod_{i=1}^n |G_i| = |G|^{n-1} |\bigcap_{j \in \mathsf{N}} G_j|$. In the case of $n=2$, it is equivalent to that $G = G_1 \circ G_2$ where $\circ$ is the binary operator of the group $G$ and $G_1 \circ G_2 \triangleq \{a \circ b : a \in G_1, b \in G_2\}$.*

### III. ABELIAN AND LINEAR GROUP CHARACTERIZATIONS

As mentioned in the introduction, the determination of capacity region for linear network codes is directly related to determining the set of entropy functions induced by abelian groups( and vector spaces). We will first study properties of abelian group characterizable entropy functions.

#### A. Abelian group characterications

**Definition 2.** *A group characterization $(G, G_1, G_2, \cdots, G_n)$ is called abelian if $G$ is an abelian group.*

Random variables (and their entropy functions) that are induced by abelian group characterizations have many nice properties.

Let $\mathbf{g} \in \mathcal{F}_n$ and $\beta$ be a non-empty subset of $\mathsf{N}$. Define $\triangle(\mathbf{g}, \beta)$ as a function in $\mathcal{F}_n$ such that for any nonempty subset $\alpha$ of $\mathsf{N}$, $\triangle(\mathbf{g}, \beta)(\alpha) = \mathbf{g}(\alpha)$ if $\alpha$ and $\beta$ are disjoint, and that $\triangle(\mathbf{g}, \beta)(\alpha) = \mathbf{g} - \sum_{j \in \beta}(\mathbf{g}(\mathsf{N}) - \mathbf{g}(\mathsf{N}\setminus\{j\}))\mathbf{u}_j$ otherwise. Or equivalently, $\triangle(\mathbf{g}, \beta) = \mathbf{g} - \sum_{j \in \beta}(\mathbf{g}(\mathsf{N}) - \mathbf{g}(\mathsf{N}\setminus\{j\}))\mathbf{u}_j$ where $\mathbf{u}_j$ is the entropy function for random variables $X_1, \cdots, X_n$ such that $H(X_j) = 1$ and $H(X_i) = 0$ for all $i \neq j$.

**Proposition 5.** *If $\mathbf{g}$ is abelian group characterizable, then $\triangle(\mathbf{g}, \beta) \in \mathcal{F}_n$ is also abelian group characterizable.*

In other words, Proposition 5 says that if random variables $X_1, \cdots, X_n$ can be characterized by subgroups of an abelian group, then $X_1, \cdots, X_n$ can be "decomposed" into $2n$'s random variables $Y_1, \cdots, Y_n, Z_1, \cdots, Z_n$ such that

1) $X_i = (Y_i, Z_i)$ for all $i \in \mathsf{N}$;
2) $H(Y_\mathsf{N}, Z_\mathsf{N}) = H(Y_\mathsf{N}) + \sum_{i \in \mathsf{N}} H(Z_i)$;
3) $H(Y_i | Y_{\mathsf{N}\setminus i}) = 0$ for all $i \in \mathsf{N}$.

Therefore, each $X_i$ is decomposed to two random variables $Y_i$ and $Z_i$ such that $Z_i$ is independent of all other random variables, and all the "interaction" among $X_i$ is captured through $Y_i$. Consequently, to study the properties of $(X_1, \cdots, X_n)$, it is sufficient to study $(Y_1, \cdots, Y_n)$ instead.

**Lemma 1.** *Let $X$ and $Y$ be two random variables. Then we can construct a random variable $W$ such that*

1) $H(W|X) = H(W|Y) = 0$
2) *If $Z$ is another random variables such that $H(Z|X) = H(Z|Y) = 0$, then $H(Z|W) = 0$.*

*For notation simplicity, $W$ will be denoted as $W = X * Y$.*

**Remark:** $W = X * Y$ is unique in the sense that two random variables are said to be equal if one is a function of the other and vice versa.

Let $X, Y$ and $Z$ be random varbiles, then $(X*Y)*Z = X*(Y*Z)$. and $X*Y = Y*X$. In other words, the operator "*" is associative and commutative.

Roughly speaking, $X*Y$ is the random variable with the maximal entropy such that it is a function of either $X$ or $Y$ alone. Note, that the entropy of $X*Y$ is always upper bounded by the mutual information between $X$ and $Y$. In fact, the bound is not tight in general. However, as will be demonstrated in the following proposition, the bound is always tight if $(X, Y)$ has an abelian group characterization.

**Proposition 6.** *Let $(G, G_1, \cdots, G_n)$ be an abelian group characterization for $X_1, \cdots, X_n$. Let $X_{n+1} = X_\alpha * X_\beta$. Then $(G, G_1, \cdots, G_n, G_{n+1})$ is an abelian group characterization for $X_1, \cdots, X_{n+1}$ where $G_{n+1} \triangleq (\bigcap_{i \in \alpha} G_i) + (\bigcap_{i \in \beta} G_i)$. Furthermore, $I(X_\alpha; X_\beta | X_{n+1}) = 0$.*

*B. Linear group characterizations*

Vector spaces are special cases of abelian groups and have many nice properties. For example, the logarithm of the size of any vector subspace is proportional to its dimension. In the following, we will show that if the group and its subgroups in a group characterization are vector space, then the induced entropy function can be defined in an alternative way by using orthogonal complements.

**Definition 3.** *Let $W_1, W_2, \cdots, W_n$ be vector subspaces of $\mathbb{F}^m$ where $\mathbb{F}$ is a finite field. Then the group characterization $(\mathbb{F}^m, W_1, W_2, \cdots, W_n)$ is called linear (with respect to the field $\mathbb{F}$).*

It is easy to verify that if $(\mathbb{F}^m, W_1, W_2, \cdots, W_n)$ is a linear group characterization for an entropy function $\mathbf{g}$, then

$$\mathbf{g}(\alpha) = \log \frac{|\mathbb{F}^m|}{|\bigcap_{i \in \alpha} W_i|} = (\dim \mathbb{F}^m - \dim W_\alpha) \log |\mathbb{F}| \quad (1)$$

Let $\mathbf{x}=(a_1, a_2, \cdots, a_m)$ and $\mathbf{y}=(b_1, b_2, \cdots, b_m)$. Define a linear operator $\langle \ \rangle : \mathbb{F}^m \times \mathbb{F}^m \to \mathbb{F}$ by $\langle \mathbf{x}, \mathbf{y} \rangle \triangleq a_1 b_2 + a_2 b_2 + \cdots + a_m b_m$. Then the vector $\mathbf{x}$ is called orthogonal to $\mathbf{y}$ (denoted as $\mathbf{x} \perp \mathbf{y}$) if and only if $\langle \mathbf{x}, \mathbf{y} \rangle = 0$. Let $W$ be a vector subspace of $\mathbb{F}^m$. Define $W^\perp = \{\mathbf{x} \in \mathbb{F}^m : \langle \mathbf{x}, \mathbf{y} \rangle = 0 \ \forall \mathbf{y} \in W\}$. Then $W^\perp$ is a vector space, and is called the orthogonal complement of $W$ (with respect to $\mathbb{F}^m$).

**Theorem 2.** *Suppose $g \in \mathcal{F}_n$ has a linear group characterization $(\mathbb{F}^m, W_1, W_2, \cdots, W_n)$. Then*

$$\mathbf{g}(\alpha) = (\dim \mathbb{F}^m - \dim W_\alpha) \log |\mathbb{F}| = \log |\mathbb{F}| \dim \sum_{i \in \alpha} W_i^\perp. \quad (2)$$

In plain words, Theorem 2 said that $\mathbf{g}$ has a linear group characterization $(\mathbb{F}^m, W_1, W_2, \cdots, W_n)$ if and only if it is proportional to the rank function for the set of vector spaces $(W_1^\perp, W_2^\perp, \cdots, W_n^\perp)$.

The study of rank functions of vector spaces has a long history. The investigation on linear dependence (which can be determined by the rank functions) led to the development of matroid theory founded by Whitney in 1935. It turns out that matroid theory plays an important role in graph theory, projective geometry, switching theory and other. The importance of Theorem 2 is that it demonstrates the equivalence of linear group characterizable entropy functions and rank functions for vector subspaces, hence creates a link between group entropy functions and matroid theory.

The following theorem is one important result that explicitly demonstrates the difference between an arbitrary group–characterizable entropy function and one induced by by abelian groups.

**Theorem 3** (Ingleton's inequality). *Let a function $\mathbf{g} \in \mathcal{F}_4$ be abelian group characterizable. Then*

$$\begin{array}{ll} \mathbf{g}(12) + \mathbf{g}(13) + \mathbf{g}(14) & \mathbf{g}(1) + \mathbf{g}(2) + \mathbf{g}(34) \\ +\mathbf{g}(23) + \mathbf{g}(24) \quad \geq & +\mathbf{g}(123) + \mathbf{g}(124) \end{array}$$

A special case of Theorem 3 was first given in [4] when $\mathbf{g}$ is a rank function for vector subspaces. By Theorem 2, Ingleton's inequality must be satisfied by all entropy functions that have a linear group characterization. It was later proved in [3] that the inequality is also satisfied by abelian group characterizable entropy functions. Using a similar proof technique, we can show that Ingleton's inequality in fact is satisfied by a more general class of entropy functions what we call "pseudo–abelian group characterizable".

**Definition 4.** *An entropy function of a set of random variables $(X_1, \cdots, X_n)$ is called pseudo-abelian group characterizable if for any disjoint subsets $\alpha$ and $\beta$ of $\mathsf{N}$,*
1) *$(X_i : i \in \alpha)$ and $(X_j : j \in \beta)$ are conditional independent given $(X_i : i \in \alpha)*(X_j : j \in \beta)$;*
2) *$(X_i : i \in \alpha)^*$ and $(X_j : j \in \beta)^*$ are conditional independent given $(X_i : i \in \alpha)^**(X_j : j \in \beta)^*$ where $(X_i : i \in \alpha)^*$ is defined as $X_{i_1}*X_{i_2}*\cdots$ for all $i_l \in \alpha$.*

**Theorem 4.** *If entropy function $\mathbf{g}$ is pseudo–abelian group characterizable, then it satisfies the Ingleton's inequality.*

This theorem is the first result showing that Ingleton's inequality has an information theoretic interpretation without resorting to the use of any group notations.

IV. BOUNDING SETS OF ENTROPY FUNCTIONS

**Definition 5.** *Define*

$$\begin{aligned} \Upsilon_n &= \{\mathbf{g} \in \mathcal{F}_n : \mathbf{g} \text{ is group characterizable. }\} \\ \Upsilon_n^{ab} &= \{\mathbf{g} \in \mathcal{F}_n : \mathbf{g} \text{ is abelian group characterizable. }\} \\ \Upsilon_n^{li} &= \{\mathbf{g} \in \mathcal{F}_n : \mathbf{g} \text{ is linear group characterizable. }\} \end{aligned}$$

As mentioned in Section 1, information inequalities and LP bounds are useful tools to determine capacity region. Due to the difficulty in characterizing the above sets of group characterizable entropy functions, capacity region bounds are usually loose. In this section, we devise a systematic approach to obtain bounds for these group characterizable entropy functions, so as to tighten the capacity region bounds.

Roughly speaking, the idea is by extension and projection. Suppose we want to obtain bounds for $\Upsilon_n$. We first show that any functions in $\Upsilon_n$ can be extended to a function in $\Upsilon_m \cap \mathcal{H}$ (with $m > n$) where $\mathcal{H}$ is usually a subset in $\mathcal{F}_m$ and has a simple explicit characterization. Then any outer bound $\mathcal{W}$ for $\Upsilon_m$ gives an outer bound for $\Upsilon_n$ by projecting the set $\mathcal{W} \cap \mathcal{H}$ back to $\Upsilon_n$. It turns out that a loose outer bound for $\Upsilon_m$ may lead to a tight outer bound for $\Upsilon_n$.

### A. Adhesive and Join extensions

**Theorem 5** (Adhesive extension). *Let $\mathbf{g} \in \Gamma_n^*$ and $\alpha$ be a subset of $\mathsf{N}$. Then there exists a function $\mathbf{h} \in \Gamma_{2n}^*$ extending $\mathbf{g}$ such that*

1) *for any $i \in \alpha$, $\mathbf{h}(i, n+i) = \mathbf{h}(i) = \mathbf{h}(n+i)$*
2) *for any nonempty subset $\beta$ of $\mathsf{N}$, $\mathbf{h}(\{i + n : i \in \beta\}) = \mathbf{h}(\{i : i \in \beta\}) = \mathbf{g}(\beta)$. Hence, $\mathbf{h}$ is an extension of $\mathbf{g}$;*
3) $\mathbf{h}(\{1, \cdots, n\}) + \mathbf{h}(\{n+1, \cdots, 2n\}) = \mathbf{h}(\{1, \cdots, 2n\}) + \mathbf{h}(\alpha)$.

*Similarly, if $\mathbf{g} \in \Upsilon_n^{ab}$, then there exists a function $\mathbf{h} \in \Upsilon_{2n}^{ab}$ extending $\mathbf{g}$ which satisfies the above three conditions.*

**Proof:** Let $\mathbf{g}$ be the entropy functions of $X_1, \cdots, X_n$. We define random variables $X_{n+1}, \cdots, X_{2n}$ such that (1) $X_{n+i} = X_i$ for $i \in \alpha$, (2) the marginal distribution of $X_{n+1}, \cdots, X_{2n}$ is the same as $X_1, \cdots, X_n$ and (3) $(X_1, \cdots, X_n)$ and $(X_{n+1}, \cdots, X_{2n})$ are conditionally independent given $(X_i : i \in \alpha)$. It can then be shown that the entropy function for $X_1, \cdots, X_{2n}$ is the desired extension of $\mathbf{g}$. Hence, the adhesive extension of $\mathbf{g}$ is closed in $\Gamma_n^*$. A similar proof can also be found in [6].

The proof for that adhesive extension is closed in $\Upsilon_n^{ab}$ will be given in the appendix. ∎

If the entropy function is abelian group characterizable, the following proposition gives another way to extension.

**Proposition 7** (Join extension). *Let $\mathbf{g} \in \Upsilon_n^{ab}$. For any nonempty disjoint subsets $\alpha, \beta$ of $\mathsf{N}$, there exists a function $\mathbf{h} \in \Upsilon_{n+1}^{ab}$ extending $\mathbf{g}$ such that (1) $\mathbf{h}(n+1, \alpha) = \mathbf{g}(\alpha) = \mathbf{h}(\alpha)$ and $\mathbf{h}(n+1, \beta) = \mathbf{g}(\beta) = \mathbf{h}(\beta)$ and (2) $\mathbf{h}(n+1, \alpha) + \mathbf{h}(n+1, \beta) = \mathbf{h}(n+1) + \mathbf{h}(\alpha, \beta)$.*

**Proof:** A direct consequence of Proposition 6. ∎

### B. Extension by coding theorems

In the previous section, extensions are made based on properties of random variables. This subsection proposed another way to extend entropy functions by using many existing coding theorems. For example, the following theorem is based on the Slepian-Wolf distributed source coding theorem.

**Theorem 6** (SW Extension). *Let $\mathbf{g} \in \Gamma_n^*$ be an entropy function for random variables $(X_1, \cdots, X_n)$. Then there exists a function $\mathbf{h} \in \overline{\Gamma}_{n+1}^*$ extending $\mathbf{g}$ such that (1) $\mathbf{h}(n+1, \alpha) = \mathbf{g}(\alpha) = \mathbf{h}(\alpha)$, (2) $\mathbf{h}(n+1) = \mathbf{g}(\alpha, \beta) - \mathbf{g}(\beta)$ and that (3) $\mathbf{h}(n+1, \beta) = \mathbf{g}(\alpha, \beta)$. Furthermore, if $\mathbf{g} \in \Upsilon_n^{li}$, then there exists $\mathbf{h} \in \Upsilon_{n+1}^{li}$ extending $\mathbf{g}$ and satisfying the three conditions.*

**Proof:** Consider two distributed sources represented by correlated random variables $X_\alpha$ and $X_\beta$. Then one can construct a message variable $X_{n+1}$ such that (1) $X_{n+1}$ is a function of $X_\alpha$, (2) the entropy of $X_{n+1}$ is about $H(X_\alpha | X_\beta)$ and (3) $X_\alpha$ can be recovered from $X_{n+1}$ and $X_\beta$ with small error probability. Using Fano's inequality, it implies that there exists $\mathbf{h} \in \overline{\Gamma}_{n+1}^*$ extending $\mathbf{g}$ and meeting the three requirements.

Now suppose that $\mathbf{g} \in \Upsilon_n^{li}$ with a linear group characterization $(\mathbb{F}^m, W_1, \cdots, W_n)$. Then for any nonempty subsets $\alpha, \beta$ of $\mathsf{N}$, there exists vector subspaces $W_{n+1}$ containing $W_\alpha$ such that (1) $W_\alpha \cap W_\beta = W_{n+1} \cap W_\beta$ and (2) $W_{n+1} + W_\beta = \mathbb{F}^m$. It is straightforward to verify that the entropy function induced by $(\mathbb{F}^m, W_1, \cdots, W_n, W_{n+1})$ is the desired extension. ∎

**Theorem 7.** *Let $\mathbf{g} \in \Gamma_n^*$ such that there exists disjoint subsets $\alpha, \beta, \gamma$ of $\mathsf{N}$ where $\mathbf{g}(\alpha, \gamma) + \mathbf{g}(\gamma) = \mathbf{g}(\alpha, \gamma) + \mathbf{g}(\beta, \gamma)$. Then there exists a function $\mathbf{h} \in \overline{\Gamma}_{n+2}^*$ extending $\mathbf{g}$ such that*

1) $\mathbf{h}(n+1, \alpha) = \mathbf{g}(\alpha) = \mathbf{h}(\alpha)$ *and* $\mathbf{h}(n+2, \gamma) = \mathbf{g}(\gamma) = \mathbf{h}(\gamma)$;
2) $\mathbf{h}(n+1) = \mathbf{g}(\alpha, \gamma) - \mathbf{g}(\gamma)$ *and* $\mathbf{h}(n+2) = \mathbf{g}(\beta) + \mathbf{g}(\gamma) - \mathbf{g}(\beta, \gamma)$;
3) $\mathbf{h}(n+1, n+2, \alpha) = \mathbf{h}(n+1, n+2)$.

**Proof:** The extension proposed in this theorem is based on the distributive source coding problem with side–information. Consider two correlated sources represented by random variables $X_\alpha$ and $X_\gamma$. Suppose $X_\beta$ is another random variable such that $X_\alpha \to X_\gamma \to X_\beta$ forms a Markov chain.

Then one can construct two message variable $X_{n+1}$ such that (1) $X_{n+1}$ is a function of $X_\alpha$ and $X_{n+2}$ is a function of $X_\gamma$, (2) the entropy of $X_{n+1}$ is about $H(X_\alpha | X_\beta)$ and the entropy of $X_{n+2}$ is about $I(X_\gamma; X_\beta)$, and (3) that $X_\alpha$ can be recovered from $X_{n+1}$ and $X_{n+2}$ with small error probability. Using Fano's inequality, it implies that there exists $\mathbf{h} \in \overline{\Gamma}_{n+1}^*$ extending $\mathbf{g}$ and meeting the three requirements. ∎

### C. Bounding by extension

We now begin with an example to illustrate the idea how to bound entropy functions by extension.

**Example 1.** *Let $\mathbf{g}$ be an abelian group characterizable entropy function for $X_1$ and $X_2$. By Proposition 6, there exists a random variable $X_3 \triangleq X_1 * X_2$ such that $I(X_1; X_2 | X_3) = 0$. Let $\mathbf{h}$ be the entropy function for $X_1, X_2, X_3$. Then $\mathbf{h}$ is an extension of $\mathbf{g}$ such that*

$$\mathbf{h}(1) = \mathbf{g}(1) \quad \mathbf{h}(2) = \mathbf{g}(2) \quad \mathbf{h}(1,2) = \mathbf{g}(1,2)$$
$$\mathbf{h}(1,3) = \mathbf{g}(1) \quad \mathbf{h}(1) + \mathbf{h}(2) = \mathbf{h}(3) + \mathbf{h}(1,2)$$

*In other words, for any $\mathbf{g} \in \Upsilon_2^{ab}$, we can extend $\mathbf{g}$ to $\mathbf{h} \in \Upsilon_3^{ab}$ such that (1) $\mathbf{h}(1) + \mathbf{h}(2) = \mathbf{h}(3) + \mathbf{h}(1,2)$, $\mathbf{h}(1,3) = \mathbf{h}(1)$ and $\mathbf{h}(2,3) = \mathbf{h}(2)$. Let*

$$\mathcal{H} = \left\{ \mathbf{h} \in \mathcal{F}_3 : \begin{array}{l} \mathbf{h}(1) + \mathbf{h}(2) = \mathbf{h}(3) + \mathbf{h}(1,2), \\ \mathbf{h}(1,3) = \mathbf{h}(1) \\ \mathbf{h}(2,3) = \mathbf{h}(2) \end{array} \right\}.$$

*Let $\mathrm{Proj}$ be the linear projection from $\mathcal{F}_3$ to $\mathcal{F}_2$. Then if $\mathcal{W}$ is an outer bound for $\Upsilon_3^{ab}$, then $\mathrm{Proj}(\mathcal{W} \cap \mathcal{H})$ gives an outer bound for $\Upsilon_2^{ab}$.*

More generally, we have the following proposition.

**Proposition 8.** *suppose that we want to bound the region $\mathcal{V}_n$. Suppose also that for any $\mathbf{g} \in \mathcal{V}_n \subseteq \mathcal{F}_n$, it can be extended to $\mathbf{h} \in \mathcal{V}_m$ and that $\mathbf{h}$ is contained in a subset $\mathcal{H}$.*

*Let $\mathrm{Proj}$ be the linear projection from $\mathcal{F}_m$ to $\mathcal{F}_n$. If $\mathcal{W}$ is an outer bound for $\mathcal{V}_m$, then $\mathrm{Proj}^{-1}(\mathcal{W} \cap \mathcal{H})$ gives an outer bound for $\mathcal{V}_n$. Furthermore, if $\mathcal{W}$ and $\mathcal{H}$ are polyhedral cone, then the obtained bound for $\mathcal{V}_n$ is also a polyhedral cone.*

In fact, the proposed method not only can find outer bound for entropy functions, it can also be used to find outer bound for entropy functions satisfying additional constraints (e.g., independency or functional dependency constraints). For example, based on the extension obtained in Theorem 7, one can obtain an outer bound for the set of entropy functions in $\Gamma_n^*$ such that $\mathbf{g}(\alpha, \gamma) + \mathbf{g}(\gamma) = \mathbf{g}(\alpha, \gamma) + \mathbf{g}(\beta, \gamma)$.

## V. CONCLUSION

In this paper, we consider random variables (and their corresponding entropy functions) that are induced by groups. Various properties for these entropy functions, especially those generated from abelian groups, are derived. By extending the entropy function to a "higher dimensional" ones, we obtain outer bounds for the set of entropy functions that are of interest.

It can be shown that the new non–Shannon information inequalities obtained in [8] and Ingleton's inequality for abelian groups induce entropy functions and can be obtained through the use of the method of extension. We believe that by making use of many source coding theorems in the literature, the method of extension can be used to derive new information inequalities and hence to tighten the bounds for a set of entropy functions of interest. As a result, a tighter capacity region for network codes can be obtained.

## APPENDIX: PROOF OF THEOREM 5

Suppose the set of random variables $(X_1, \cdots, X_n)$ has a group characterization $(G, G_1, \cdots, G_n)$. Let $\alpha$ be a nonempty subset of $\mathsf{N}$. To prove Theorem 5, it suffices to prove the following stronger statement- if $G_\alpha$ is a normal subgroup of $G$, then there exists a group-characterizabtion for random variables $(Y_1, \cdots, Y_n, Z_1, \cdots, Z_n)$ such that

1) for any nonempty subset $\beta$ of $\mathsf{N}$, $H(X_\beta) = H(Y_\beta) = H(Z_\beta)$;
2) for any $i \in \alpha$, $H(Y_i|Z_i) = H(Z_i|Y_i) = 0$. Hence, $(Y_i, i \in \alpha) = (Z_i, i \in \alpha)$;
3) $Y_\mathsf{N}$ is conditional independent of $Z_\mathsf{N}$ given $Y_\alpha$

To prove the statement, we first define the set $K$ as $\{(a,b) \in G \times G : aG_\alpha = bG_\alpha\}$. By direct counting, the cardinality of $K$ is equal to $|G||G_\alpha|$.

For any $(a,b), (c,d) \in K$, we define $(a,b) \circ (c,d)$ as the element in $(ac, bd)$. It can be verified easily that $acK = bdK$ and hence the binary operation is closed in $K$. In fact, $(K, \circ)$ is indeed a group – (1) the associativity of "$\circ$" follows from that $G$ is a group, (2) the identity element is $(e,e)$ and (3) the inverse of $(a,b) \in K$ is $(a^{-1}, b^{-1})$.

Define a group homomorphism $T_1 : K \longrightarrow G$ such that $T_1(a,b) = a$. Similarly, define $T_2(a,b) = b$. For any subgroup $L$ in $G$, we define subgroups $L^1$ and $L^2$ in $K$ as follows:

$$L^1 \triangleq \{(a,b) \in K : a \in L\} \quad = \quad T_1^{-1}(L) \quad (3)$$

$$L^2 \triangleq \{(a,b) \in K : b \in L\} \quad = \quad T_2^{-1}(L) \quad (4)$$

The subgroups $L^1$ and $L^2$ have many interesting properties including the following,

P1: $|K|/|L^1| = |K|/|L^2| = |G|/|L|$. (*This property follows from that $|K| = |G||G_\alpha|$ and $|L_1| = |L_2| = |L||G_\alpha|$.*)
P2: for all $j \in \alpha$, $G_j^1 = G_j^2$ (*Or equivalently, $T_1^{-1}(G_j) = T_2^{-1}(G_j)$*);
P3: For any subgroup $H$ in $G$ and $i = 1, 2$, $H^i \cap G_\alpha^i = (H \cap G_\alpha)^i$ (i.e., $T_i^{-1}(H) \cap T_i^{-1}(G_\alpha) = T_i^{-1}(H \cap G_\alpha)$). Hence, $|H^i \cap G_\alpha^i| = |H \cap G_\alpha||G_\alpha|$.
P4: For any subgroups $H$ and $L$ in $G$,

$$H^1 \cap G_\alpha^1 \cap L^2 = \{(a,b) : a \in H \cap G_\alpha, b \in L \cap G_\alpha\}.$$

Hence, $|H^1 \cap G_\alpha^1 \cap L^2| = |H \bigcap G_\alpha||L \bigcap G_\alpha|$.

Now, we define subgroups $H_i \triangleq G_i^1 = T_1^{-1}(G_i)$ and $L_i \triangleq G_i^2 = T_2^{-1}(G_i)$ for $i \in \mathsf{N}$. Let $(Y_1, \cdots, Y_n, Z_1, \cdots, Z_n)$ be random variables induced by the group characterization $(K, H_1, \cdots, H_n, L_1, \cdots, L_n)$. Then by property 3, for any nonempty subset $\beta$ of $\mathsf{N}$, $H(X_\beta) = H(Y_\beta) = H(Z_\beta)$. By property 2, for any $i \in \alpha$, $H(Y_i|Z_i) = H(Z_i|Y_i) = 0$. Hence, $(Y_i, i \in \alpha) = (Z_i, i \in \alpha)$. Finally, by properties 1 and 4,

$$\frac{|K|}{|H_\mathsf{N}|} \frac{|K|}{|L_\mathsf{N}|} = \frac{|G|}{|G_N|} \frac{|G|}{|G_N|} \quad (5)$$

$$= \frac{|G|}{|G_\alpha|} \frac{|G||G_\alpha|}{|G_N||G_N|} \quad (6)$$

$$= \frac{|K|}{|H_\alpha|} \frac{|K|}{|H_\mathsf{N} \cap L_\mathsf{N}|} \quad (7)$$

Hence, $Y_\mathsf{N}$ is conditional independent of $Z_\mathsf{N}$ given $Y_\alpha$ and $(K, H_1, \cdots, H_n, L_1, \cdots, L_n)$ is the desired group characterization.

## REFERENCES

[1] T.H. Chan, "Capacity regions for linear and abelian network codes," *Third Workshop on Network Coding, Theory, and Applications* 2007, San Diego, California
[2] T. H. Chan, R. W. Yeung, "On a relation between information inequalities and group theory," *IEEE Trans. on Inform. Theory*, vol. 48, pp. 1992-1995, July 2002.
[3] T. H. Chan, "Aspects of Information inequalities and its applications," M.Phil. Thesis, Dept. of Information Engineering, The Chinese University of Hong Kong, September 1998.
[4] A.W. Ingleton, "Characterization of Matroids," Combinatorial mathematics and it's application. Academic Press, 1971.
[5] F. Matúš, "Probabilistic conditional independence structures and matroid theory: Background", International Journal of General Systems, vol. 22, pp. 185-196, 1994.
[6] F. Matus, "Inequalities for Shannon entropies and adhesivity of polymatroids," *Proceedings CWIT 2005*, Montreal, Canada, pp.28-31
[7] R. W. Yeung, "A framework for linear information inequality," IEEE Trans. of Information Theory, vol. 43, pp. 1924-1934, November, 1997.
[8] Z. Zhang and R. W. Yeung, " On the characterization of entropy function via information inequalities," IEEE Transactions of Information Theory, vol. 44, pp. 1440-1452, 1998.